\documentclass[aps,prb,twocolumn,superscriptaddress,longbibliography,floatfix]{revtex4-1}

\usepackage[T1]{fontenc}
\usepackage[utf8]{inputenc}
\usepackage{lmodern}
\usepackage{microtype}
\usepackage{amsmath,amssymb,bm}
\usepackage{graphicx}
\usepackage{booktabs}
\usepackage{siunitx}
\usepackage{xcolor}
\usepackage{hyperref}
\usepackage{comment}

\newcommand{\kk}{\bm{k}}
\newcommand{\qq}{\bm{q}}
\newcommand{\rr}{\bm{r}}
\newcommand{\RR}{\bm{R}}
\newcommand{\QQ}{\bm{Q}}
\newcommand{\MM}{\bm{M}}
\newcommand{\ww}{\omega}
\newcommand{\ii}{\mathrm{i}}
\newcommand{\BZ}{\mathrm{BZ}}
\newcommand{\Tr}{\mathrm{Tr}}

\begin{document}

\title{Effects of gold cluster intercalation in graphene: \\stationary waves and modified QPI features}

\author{Cristina Bena}
\affiliation{Universit\'e Paris Saclay, CNRS, CEA, Institut de Physique Th\'eorique, 91191, Gif-sur-Yvette, France}
\author{Marion Cranney}
\affiliation{Institut de Sciences des Mat\'eriaux de Mulhouse, CNRS-UMR 7361
Universit\'e de Haute Alsace, Mulhouse 68093, France}
\author{Poonam Kumari}
\affiliation{Universit\'e Paris Saclay, CNRS, CEA, Institut de Physique Th\'eorique, 91191, Gif-sur-Yvette, France}
\author{Alberto Zobelli}
\affiliation{Universit\'e Paris Saclay, CNRS, Laboratoire de Physique des Solides, Orsay, France}
\author{Laurent Simon}
\affiliation{Institut de Sciences des Mat\'eriaux de Mulhouse, CNRS-UMR 7361
Universit\'e de Haute Alsace, Mulhouse 68093, France}
\date{\today}

\begin{abstract}
Gold intercalation beneath epitaxial graphene on SiC produces a cluster phase with unusual standing waves and quasiparticle-interference (QPI) features concentrated near the graphene $M$ points\cite{CranneyEPL2010}. We show that this can be explained by Au intercalation below graphene hollow sites, which induces a local scattering potential on the six surrounding carbon atoms. Within a $T$-matrix treatment, this ring-like scatterer produces elliptical QPI structures centered near $M$, in agreement with the experimental FT-STS measurements. We further show that these QPI features naturally generate the nearly stationary standing-wave patterns observed in real space. Finally, we compute the local-density-of-states contrast on and off a small cluster and show that its sign and magnitude are strongly energy dependent, consistent with the experimental observations.
\end{abstract}

\maketitle

\section{Introduction}
\label{sec:intro}

Intercalation provides a direct chemical route for modifying the electronic structure of epitaxial graphene. In graphene on SiC, intercalants may reside between the graphene sheet and the buffer layer, or penetrate below the buffer layer and decouple it from the substrate. These configurations can produce quasi-free-standing graphene, ordered superstructures, dense interfacial metallic layers, or sparse cluster phases. The corresponding electronic response is often much richer than a rigid shift of the Dirac point: intercalants can generate new bands, modify the effective band structure, redistribute spectral weight, and strongly reshape the Van-Hove region where the density of states is large \cite{CranneyEPL2010,RiedlPRL2009,GierzPRB2010,FortiPRB2011,BriggsNanoscale2019,PremlalAPL2009,NairPRB2012,Nair2016AuGrapheneSiC,RosenzweigPRB2019VHS,RosenzweigPRL2020Overdoping,LinkPRB2019Gd}.
Rare-earth and alkali-metal intercalants provide particularly clear
examples: Yb/K and Er intercalation can drive graphene to the
Lifshitz-transition regime, while Gd intercalation produces strong
many-body renormalization near the Van-Hove singularity (VHS).
\cite{RosenzweigPRL2020Overdoping,LinkPRB2019Gd,ZaarourPRR2023}.

Gold-intercalated graphene is a particularly useful example. Scanning tunnelling microscopy (STM) studies revealed several Au-induced phases under monolayer graphene on SiC, including a disordered cluster phase known as the ``ostrich-leather'' phase \cite{PremlalAPL2009}. In this phase, intercalated Au atoms form a network of nanoscale clusters. Fourier-transform scanning tunnelling spectroscopy (FT-STS) measurements display strong features around the graphene $M$ points, together with real-space standing waves whose period remains almost constant over a finite bias window \cite{CranneyEPL2010}. These patterns differ from the usual QPI of weakly perturbed graphene, where the dominant features are tied to intravalley and intervalley scattering between constant-energy contours around $K$ and $K'$ \cite{RutterScience2007,BrihuegaPRL2008,BenaPRL2008}.

In a recent work \cite{kinkpaper} we developed an effective microscopic description of the cluster phase which allowed us to conclude that the Au atoms sit underneath the graphene hollow sites and couple to the six neighbouring carbon atoms either through a local electrostatic potential or through Au--C hybridization.  The purpose of the present work is to investigate the consequences of such an impurity for the QPI patterns and compare the results with the experimental results of Ref.~\onlinecite{CranneyEPL2010}.

Our first observation is that the sixth-nearest-neighbour graphene model, which accurately reproduces the DFT dispersion, produces strongly warped constant-energy contours at large positive energies, and an extended upper VHS. When this spectral structure is combined with impurity ring-like scattering and treated in the $T$-matrix formalism, the resulting Fourier maps exhibit elliptic $M$-centered structures similar to those observed experimentally. Using a minimal real-space model in which the QPI signal is represented by double-lobe elliptical features centered around the $M$ points, we show that such features naturally generate the observed standing waves. The finite width of the lobes and the distance between them control the coherence length of the real-space pattern. The main ingredient responsible for the elliptical QPI structures is the ring geometry of the Au--C coupling.

We also calculate the spatial dependence of the LDOS inside and near an individual cluster and show that the contrast between the LDOS inside and outside the cluster depends strongly on energy, in qualitative agreement with the experimental observations.

The paper is organized as follows. In Sec.~\ref{sec:pristine} we introduce the pristine-graphene model and calculate the spectral function at various energies. In Sec.~\ref{sec:impurity} we describe the ring-impurity model, we calculate the corresponding QPI features, and we connect them to the formation of real-space standing waves. In Sec.~\ref{sec:realspace} we calculate the LDOS as a function of energy and position and we compare it to the experimental observations. We conclude in Sec.~\ref{sec:concl}.

\section{Pristine graphene}
\label{sec:pristine}

We use the same graphene reference model as in Ref.~\onlinecite{kinkpaper}. The graphene $p_z$ Hamiltonian is written in the sublattice basis
\begin{equation}
  \Psi_{\kk}=\begin{pmatrix}c_{A\kk}\\ c_{B\kk}\end{pmatrix},
\end{equation}
with
\begin{equation}
  H_0(\kk)=
  \begin{pmatrix}
    \epsilon_{\rm same}(\kk)-\mu & h_{\rm AB}(\kk) \\
    h_{\rm AB}^*(\kk) & \epsilon_{\rm same}(\kk)-\mu
  \end{pmatrix} .
  \label{eq:H0}
\end{equation}
The off-diagonal and same-sublattice terms are retained up to sixth-nearest neighbours,
\begin{align}
  h_{\rm AB}(\kk)
  &= t_1 f_1(\kk)+t_3 f_3(\kk)+t_4 f_4(\kk),
  \label{eq:hab}\\
  \epsilon_{\rm same}(\kk)
  &= t_2 g_2(\kk)+t_5 g_5(\kk)+t_6 g_6(\kk).
  \label{eq:epssame}
\end{align}
Here $f_n$ denotes the structure factor for the $n$th opposite-sublattice neighbour shell, while $g_n$ denotes the corresponding same-sublattice structure factor:
\begin{align}
  f_n({\bf k})
  &=
  \sum_{\bm{\delta}\in S_n^{AB}}
  e^{i{\bf k}\cdot\bm{\delta}},
  \qquad n=1,3,4,
  \label{eq:fnAB}
  \\
  g_n({\bf k})
  &=
  \sum_{\bm{\rho}\in S_n^{AA}}
  e^{i{\bf k}\cdot\bm{\rho}},
  \qquad n=2,5,6,
  \label{eq:gnAA}
\end{align}
where
\begin{align}
  S_1^{AB}
  &=\{\boldsymbol{\delta}_1,\boldsymbol{\delta}_2,\boldsymbol{\delta}_3\}\nonumber\\
    S_2^{AA}
  &=
  \left\{
  \pm {\bf a}_1,
  \pm {\bf a}_2,
  \pm({\bf a}_2-{\bf a}_1)
  \right\},
  \nonumber\\
  S_3^{AB}
  &=\{-2\boldsymbol{\delta}_1,-2\boldsymbol{\delta}_2,-2\boldsymbol{\delta}_3\},
\nonumber\\
  S_4^{AB}
  &=
  \left\{
  2\boldsymbol{\delta}_i-\boldsymbol{\delta}_j
  \;,\;
  i,j\in\{1,2,3\},\ i\neq j
  \right\},
  \nonumber\\
  S_5^{AA}
  &=
  \left\{
  \pm({\bf a}_1+{\bf a}_2),
  \pm(2{\bf a}_1-{\bf a}_2),
  \pm({\bf a}_1-2{\bf a}_2)
  \right\},
  \nonumber\\
  S_6^{AA}
  &=
  \left\{
  \pm2{\bf a}_1,
  \pm2{\bf a}_2,
  \pm2({\bf a}_2-{\bf a}_1)
  \right\},
  \label{eq:longerRangeNeighborShells}
\end{align} 
the graphene Bravais-lattice vectors are given by
\begin{equation}
  {\bf a}_1=\boldsymbol{\delta}_1-\boldsymbol{\delta}_2,
  \qquad
  {\bf a}_2=\boldsymbol{\delta}_1-\boldsymbol{\delta}_3,
\end{equation}
and the nearest-neighbor vectors are
\begin{equation}
  \boldsymbol{\delta}_1 = a(0,1),   \boldsymbol{\delta}_2 = a\left(\frac{\sqrt{3}}{2},-\frac{1}{2}\right),   \boldsymbol{\delta}_3 = a\left(-\frac{\sqrt{3}}{2},-\frac{1}{2}\right),
\end{equation}
where $a$ is the carbon--carbon distance. 

\begin{figure}[t]
  \centering
  \includegraphics[width=0.4\textwidth]{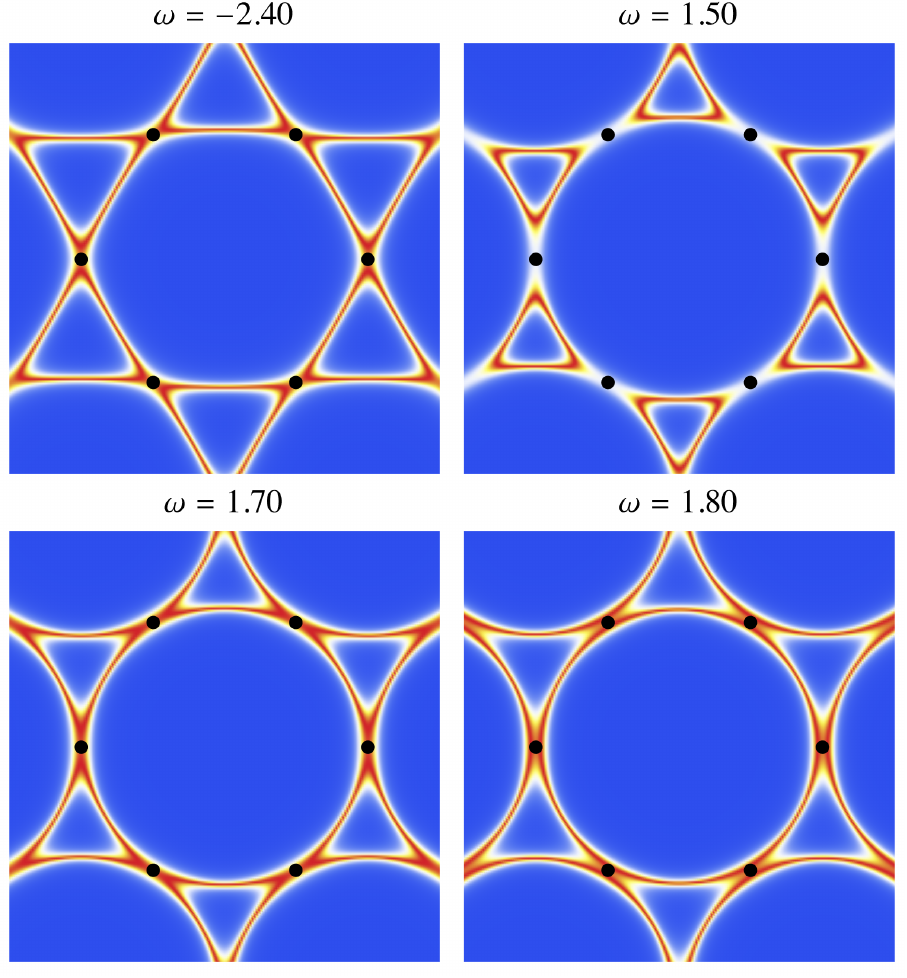}
  \caption{Pristine 6NN graphene spectral function at different energies. The black dots denote the $M$ points of the BZ.}
  \label{fig:pristine_spectral}
\end{figure}

This parametrization is important for the present problem because the upper and lower van-Hove regions are not related by particle-hole symmetry once second- through sixth-neighbour hoppings are included \cite{kinkpaper,ReichPRB2002}. In particular, the positive-energy contours near the VHS are much more strongly warped than their negative-energy counterparts, as shown in Fig.~\ref{fig:pristine_spectral}.

In the numerical calculations we use the Wannier-fitted hopping values \cite{kinkpaper,Wannier1937}
\begin{eqnarray}
  t_1 &=& -2.937\,{\rm eV}\nonumber \\
  t_2 &=& 0.249\,{\rm eV}\nonumber\\   t_3 &=& -0.260\,{\rm eV}\nonumber\\
  t_4 &=& 0.025\,{\rm eV}\nonumber\\  t_5 &=& 0.050\,{\rm eV}\nonumber\\ t_6&=&-0.024\,{\rm eV}.
  \label{eq:TBparams}
\end{eqnarray}
The retarded pristine Green's function is
\begin{equation}
  G_0(\kk,\ww)=\left[(\ww+\ii\eta)\mathbb{I}_2 -H_0(\kk)\right]^{-1},
  \label{eq:g0}
\end{equation}
with a broadening $\eta=0.2\,\mathrm{eV}$. 
We use this relatively large broadening to mimic, at the level of the effective model, the disorder and finite lifetime associated with the Au-cluster phase. As shown in Ref.~\onlinecite{kinkpaper}, the most important effect of the Au-induced local electrostatic potential, which is the dominant impurity factor at the energies considered here, is an increase in the graphene linewidth associated with an decreased quasiparticle lifetime. We have checked that the formation of M-centered ellipses is generic and valid for a wide range of $\eta \sim 0.1 - 0.3$.

The corresponding spectral function is
\begin{equation}
  A_0(\kk,\ww)= -\frac{1}{\pi}\,\rm{Im}\,\Tr_{A,B}G_{0}(\kk,\ww).
  \label{eq:A0}
\end{equation}
The chemical potential $\mu=-0.377\,\mathrm{eV}$ is chosen so that the Dirac point lies at the Fermi level. Figure~\ref{fig:pristine_spectral} shows the pristine spectral function at several energies, close to the lower and upper VHSs. While at the lower VHS the contours are close to equilateral triangles, at the upper VHS the contours are strongly warped and the spectral weight is extended along the $KK'$ direction close to the $M$ point.

This corresponds to a slight flattening of the band near the
upper VHS. Similar extended or flattened spectral features near
the graphene VHS have been observed in strongly electron-doped
intercalated graphene, including Yb-intercalated graphene further
doped by potassium adsorption, Gd-intercalated graphene, and
long-range ordered Er-intercalated graphene
\cite{RosenzweigPRB2019VHS,RosenzweigPRL2020Overdoping,LinkPRB2019Gd,ZaarourPRR2023}.

\section{Impurity effects}
\label{sec:impurity}

\subsection{Hollow-site Au impurity model}

We model an intercalated Au atom as a finite-range impurity centered below a graphene hollow site. The relevant local Hilbert space consists of the Au-derived orbitals and the six carbon $p_z$ orbitals surrounding the hollow site. Labeling the carbon sites by $j=1,\ldots,6$, their positions relative to the hollow-site center by $\rr_j$, and their sublattice indices by $s_j=A,B$, the local impurity Hamiltonian can be written as
\begin{equation}
  H_{\rm imp}\!\!
  =\!\!
  \sum_{\alpha}\varepsilon_{\alpha}d_{\alpha}^{\dagger}d_{\alpha}
  +\!\!
  \sum_{j,\alpha}
  \left(
    V_{j\alpha}c_j^{\dagger}d_{\alpha}
    +
    V_{j\alpha}^{*}d_{\alpha}^{\dagger}c_j
  \right)
  +
  U\! \sum_{j=1}^{6}c_j^{\dagger}c_j.
  \label{eq:himp}
\end{equation}
Here $d_{\alpha}$ annihilates an electron in the Au orbital $\alpha$, $V_{j\alpha}$ is its hybridization with ring site $j$, and $U$ is an effective electrostatic shift on the neighboring carbon atoms. The last term therefore also acts as a short-range elastic scattering potential with the spatial form factor of the six-carbon ring.

The local-potential and hybridization descriptions are closely related away from an Au-derived resonance. Integrating out the Au orbitals gives the energy-dependent potential in the six-site ring space
\begin{equation}
  V_{\rm eff}(\ww)
  =
  U\mathbb{I}_6
  +
  \sum_{\alpha}
  \frac{\bm v_{\alpha}\bm v_{\alpha}^{\dagger}}
  {\ww-\varepsilon_{\alpha}+i\eta_{\alpha}},
  \qquad
  [\bm v_{\alpha}]_j=V_{j\alpha}.
  \label{eq:Veff_random}
\end{equation}
Far from the corresponding hybridization energies, $\varepsilon_{\alpha}$, the second term varies only weakly over a restricted energy window and acts approximately as an additional finite-range potential on the same six surrounding carbon atoms.
Since the QPI features considered here lie far from the relevant Au hybridization levels ($\sim$ -2eV -- -3eV)\cite{kinkpaper}, we use the minimal approximation $V_{\rm eff}(\ww)\simeq U\mathbb{I}_6$; the fitted value of $U$ may then also absorb the slowly varying off-resonant contribution of the Au orbitals. In what follows we will take $U=$ 3eV, a value which is of the same order of magnitude as that of Ref.~\onlinecite{kinkpaper}.

\subsection{QPI and single-impurity $T$ matrix}
\label{subsec:tmatrix}

We compute the quasiparticle-interference signal using the standard single-impurity $T$-matrix formalism
\cite{BenaPRL2008,Kaasbjerg2020Tmatrix,Kot2020DefectsSCTMA}. Because a hollow-site Au impurity acts on a six-carbon ring rather than on a single lattice site, it is convenient first to formulate the scattering problem in this six-dimensional local space. We introduce the $2\times6$ embedding matrix
\begin{equation}
  \left(\mathcal P_{\kk}\right)_{s j}
  =
  \delta_{s,s_j}e^{i\kk\cdot\rr_j},
  \qquad s=A,B,
  \label{eq:Pembed}
\end{equation}
which projects graphene Bloch states onto the six carbon sites surrounding the impurity. The clean graphene Green's function projected onto the ring is then
\begin{equation}
  g_0(\ww)
  =
  \int_{\BZ}[d\kk]\,
  \mathcal P_{\kk}^{\dagger}
  G_0(\kk,\ww)
  \mathcal P_{\kk}.
  \label{eq:ring_green}
\end{equation}
Each matrix element of $g_0$ is the amplitude for an electron to propagate through pristine graphene between two sites of the six-carbon ring, including their sublattice character and relative phase.

The ring-space $T$ matrix satisfies the Dyson equation
$T=V_{\rm eff}+V_{\rm eff}g_0T$ and is consequently
\begin{equation}
  T(\ww)
  =
  V_{\rm eff}(\ww)
  \left[
    \mathbb{I}_6-g_0(\ww)V_{\rm eff}(\ww)
  \right]^{-1}.
  \label{eq:ring_T}
\end{equation}
This expression resums the full multiple-scattering series
$V_{\rm eff}+V_{\rm eff}g_0V_{\rm eff}+V_{\rm eff}g_0V_{\rm eff}g_0V_{\rm eff}+\cdots$.
Thus $T(\ww)$ is not the bare impurity potential, but the complete amplitude for arbitrarily many returns to, and scatterings within, the six-site impurity complex. Possible impurity resonances correspond to poles of this matrix, determined by
$\det[\mathbb{I}_6-g_0(\ww)V_{\rm eff}(\ww)]=0$.

The scattering amplitude connecting graphene Bloch states is obtained by embedding the local result back into the two-sublattice basis,
\begin{equation}
  \mathcal T(\kk,\kk';\ww)
  =
  \mathcal P_{\kk}
  T(\ww)
  \mathcal P_{\kk'}^{\dagger}.
  \label{eq:Tembed}
\end{equation}
The phase factors in $\mathcal P_{\kk}$ retain the finite size and sixfold geometry of the impurity. They generate the hollow-site form factor that distinguishes the present ring scatterer from a structureless onsite defect.

The impurity-induced correction to the Fourier-transformed local density of states is
\begin{equation}
\begin{split}
  \delta\rho(\qq,\ww)
  =
  -\frac{1}{\pi}\,{\rm Im}
  \int_{\BZ}[d\kk]\,
  \Tr_{\rm sub}
  \Big[
  G_0(\kk,\ww)
  \mathcal T(\kk,\kk+\qq;\ww)
  \\
  \times
  G_0(\kk+\qq,\ww)
  \Big].
\end{split}
  \label{eq:qpi_general}
\end{equation}
Here
 \begin{equation}
[d{\bf k}]
\equiv
\frac{d^2{\bf k}}{S_{\mathrm{BZ}}},
\qquad
S_{\mathrm{BZ}}
=
\frac{8\pi^2}{\sqrt{3}\,a^2},
\end{equation}
The trace is over the two graphene sublattices, and $\qq$ is the momentum transferred by the impurity. The first Green's function describes propagation with incoming momentum $\kk$, while the second describes propagation after scattering to $\kk+\qq$. In the numerical plots we display $|\delta\rho(\qq,\ww)|$, which is the quantity most directly comparable to the Fourier amplitude measured in FT-STS maps.

\begin{figure*}[t]
  \centering
  \includegraphics[width=0.8\textwidth]{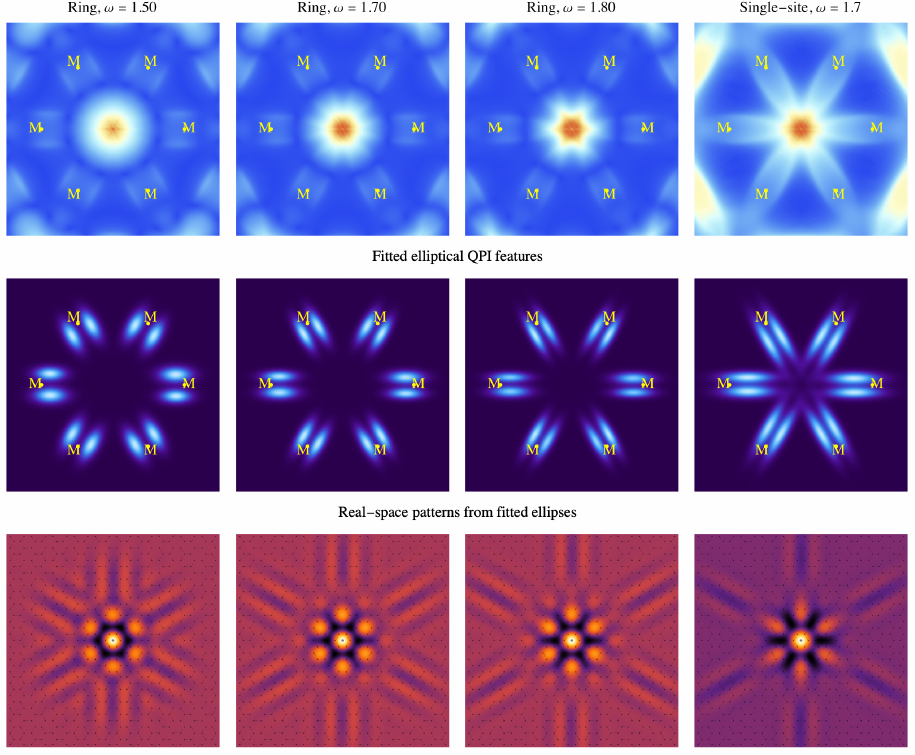}
  \caption{QPI patterns (first row), the corresponding double-lobe model (second row), and the resulting real-space standing waves (third row). We take $U=$ 3eV. Compact $M$-centered lobes give coherent nearly stationary patterns, whereas the standing waves that correspond to extended features dephase much more rapidly in real space.}
  \label{fig:standing_wave_model}
\end{figure*}

\begin{figure}[t]
  \centering
  \includegraphics[width=0.23\textwidth]{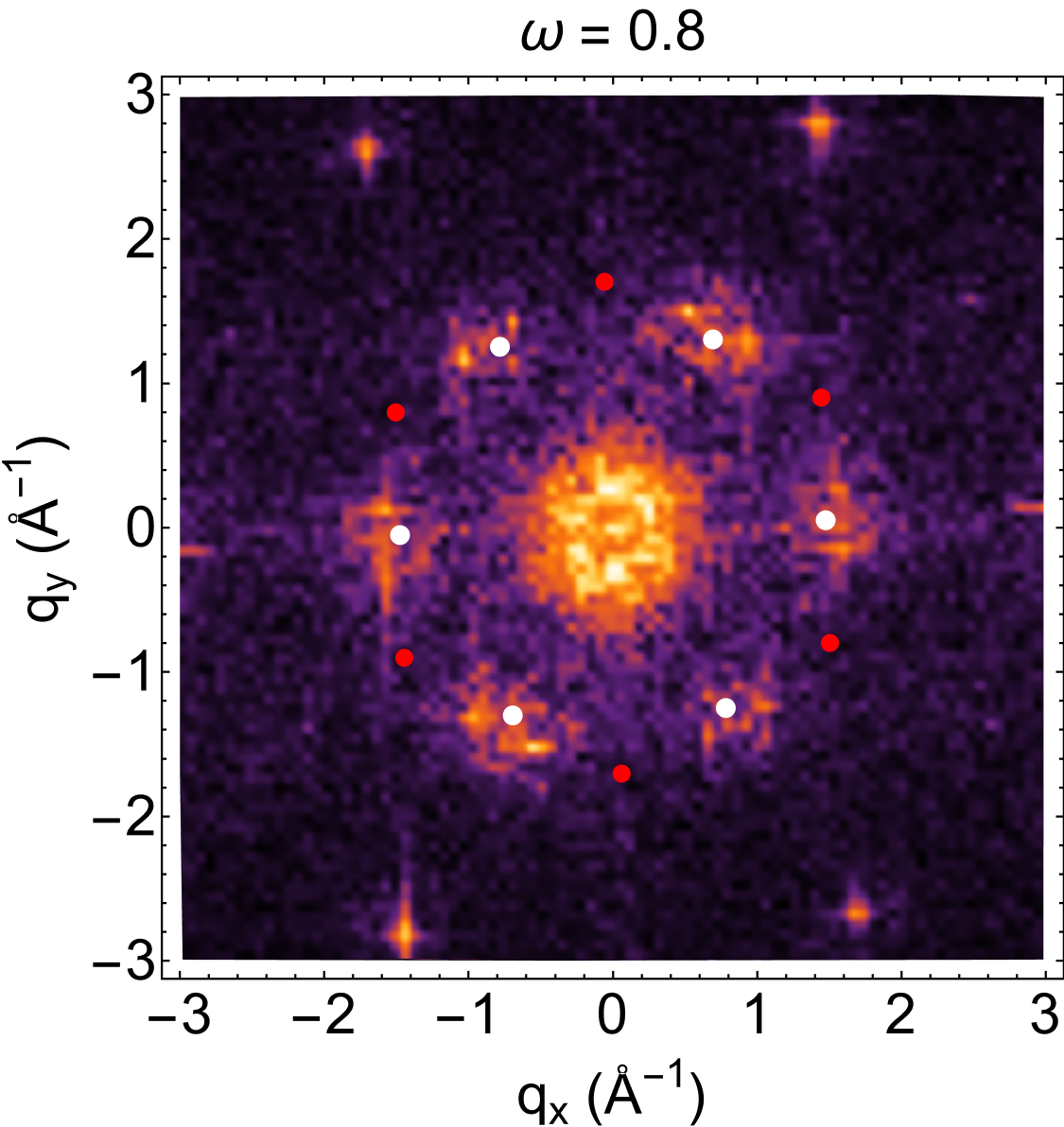}
    \includegraphics[width=0.23\textwidth]{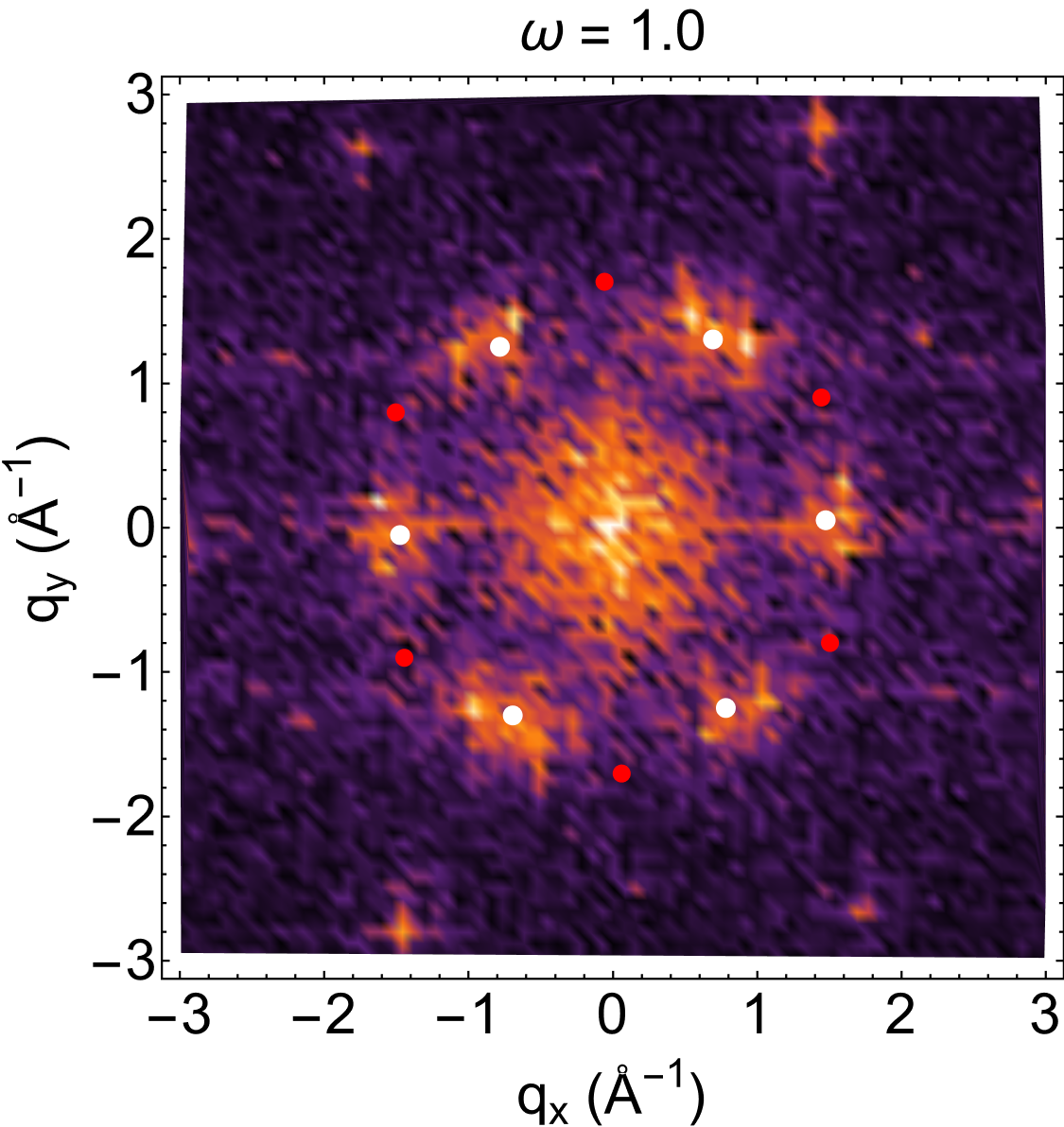}
  \caption{Experimental QPI maps of the Au-cluster phase at two energies close to the upper VHS, $\omega=0.8$ eV and $\omega=1$ eV. The red dots denote the $K$ points of the BZ, and the white dots denote the $M$ points.}
  \label{fig:exp_ftsts}
\end{figure}

The first three plots in the top row of Fig.~\ref{fig:standing_wave_model} show the QPI maps at energies close to the upper VHS, which is here at $1.74\,\mathrm{eV}$. The QPI response develops two compact lobes in the vicinity of each $M$ point.  Close to the VHS, however, their shape and size vary only weakly with energy. These two-lobe features reproduce the characteristic structure of the experimental FT-STS maps shown in Fig.~\ref{fig:exp_ftsts}.

In our analysis we choose parameters such that the Dirac point lies at the Fermi energy. The experimental measurements may be shifted, with a Dirac point below the Fermi energy, up to $-0.5\,\mathrm{eV}$. Moreover, the precise energy of the VHS is not known: other factors in the sample may influence the band structure. Thus the comparison of the theoretical and experimental intervals of the energies for which the elliptical features arise in the QPI is only qualitative. 

For comparison, the last panel of the top row of Fig.~\ref{fig:standing_wave_model}  shows the average QPI response of a localized impurity acting on single A or B carbon sites. For this impurity the QPI spectra contain extended line-like structures rather than compact double ellipses. 
We therefore identify the ring shape of the Au-induced scattering potential as the essential ingredient that produces the double-lobe elliptical QPI features.

\subsection{Standing waves from $M$-centered QPI lobes}
\label{subsec:standing_waves}

To make the connection between the calculated QPI maps and the real-space standing waves more transparent, we approximate the $M$-centered QPI features by a minimal analytic model. Around each of the six $M$ points we place two elliptic Gaussian lobes,
\begin{align}
  I_M(\qq)
  &= \sum_{\ell=1}^{6}\sum_{s=\pm}
  A \exp\biggl[-\frac{1}{2}
  \biggl(
  \frac{\bigl[(\qq-\QQ_{\ell s})\cdot\hat e_{\ell,1}\bigr]^2}{\sigma_1^2}
  \nonumber\\
  &\qquad\qquad\qquad
  +
  \frac{\bigl[(\qq-\QQ_{\ell s})\cdot\hat e_{\ell,2}\bigr]^2}{\sigma_2^2}
  \biggr)
  \biggr],
  \label{eq:qpi_gaussian_model}
\end{align}
where
\begin{equation}
  \QQ_{\ell s}=\MM_\ell+s\Delta q\,\hat n_\ell .
\end{equation}
The three geometric parameters are the lobe length, $\sigma_1$, the lobe width, $\sigma_2$, and the separation, $2\Delta q$, between the two maxima near a given $M$ point. The unit vectors $\hat e_{\ell,1}$ and $\hat e_{\ell,2}$ fix the orientation of the ellipses, while $\hat n_\ell$ fixes the direction of the splitting.

The corresponding standing-wave pattern is obtained by Fourier transforming the intensity generated by this model. For an extended field of view, it is useful to represent the lobes by a set of discrete Fourier modes sampled from Eq.~\eqref{eq:qpi_gaussian_model},
\begin{equation}
  F(\rr)=\sum_n w_n\cos(\qq_n\cdot\rr),
  \label{eq:finite_modes}
\end{equation}
with weights $w_n=I_M(\qq_n)$. 

The middle row of Figure~\ref{fig:standing_wave_model} shows the double-lobe elliptical fit.
As the energy increases from $\omega=1.5~\mathrm{eV}$ to $1.8~\mathrm{eV}$, the two
ring-impurity lobes move progressively closer to the corresponding
$M$ point, with their separation $2\Delta q$ decreasing from
approximately $0.64\mathrm{\AA}^{-1}$ to $0.42\mathrm{\AA}^{-1}$. At the same time, the lobes become
longer and narrower: the lobe length $\sigma_1$ increases from
approximately $0.33\mathrm{\AA}^{-1}$ to $0.41\mathrm{\AA}^{-1}$, whereas the lobe width $\sigma_2$
decreases from approximately $0.15\mathrm{\AA}^{-1}$ to $0.11\mathrm{\AA}^{-1}$. Consequently, the
aspect ratio $\sigma_1/\sigma_2$ increases from approximately $2.2$ to
$3.6$. By comparison, at $\omega=1.7~\mathrm{eV}$ a localized
single-sublattice impurity produces a substantially more extended
feature, with $\sigma_1\simeq 0.54\mathrm{\AA}^{-1}$, $\sigma_2\simeq 0.12\mathrm{\AA}^{-1}$, and
$\sigma_1/\sigma_2\simeq 4.4$, consistently with its more line-like
QPI response.

The bottom row of Figure~\ref{fig:standing_wave_model} shows representative real-space patterns generated from the $M$-centered double ellipses. The dominant period is fixed by wavevectors close to $M$, while the finite size and shape of the ellipses control the coherence length. The elliptical features generated by a ring impurity give rise to robust standing waves, with an apparent wavelength comparable to that observed in STM. 
By contrast, the QPI patterns generated by a single-site impurity give rise to real space patterns that dephase rapidly and do not exhibit a coherent resonator-like character.

\begin{figure}[t]
  \centering
  \includegraphics[width=0.5\textwidth]{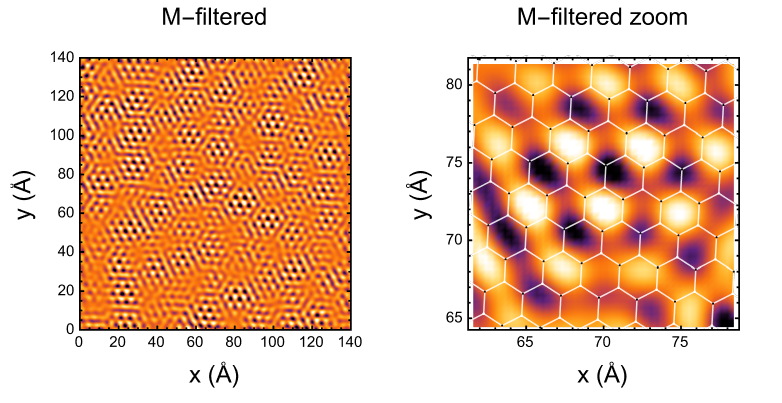}
\caption{Experimental standing-wave pattern obtained by Fourier
filtering the $dI/dV$ map measured at $E=0.8~\mathrm{V}$. The filter
retains only the six elliptical Fourier-space regions centered near the
symmetry-equivalent $M$ points, with their long axes approximately
parallel to the $M$--$K$--$\Gamma$ directions; all remaining Fourier
components are suppressed before performing the inverse Fourier
transform. }
  \label{fig:exp_standing_waves}
\end{figure}

For comparison, Fig.~\ref{fig:exp_standing_waves} shows the real-space standing-wave pattern observed experimentally in the Au-cluster phase, filtered to show only the oscillations with wavevectors in the vicinity of the M points.  Note the very good agreement with the theoretical analysis in the first three panels of the bottom row of Fig.~\ref{fig:standing_wave_model}: standing waves with a $2 \times 2$ periodicity arise and preserve spatial coherence over approximatively a period of about three oscillations.

This simple exercise explains why the cluster-phase standing waves can appear nearly stationary over a finite bias window. The relevant wavevectors are pinned by the upper-VHS region and remain close to $M$ as the energy changes.

\section{LDOS as a function of energy and position}
\label{sec:realspace}

We finally investigate the contrast between the LDOS on a cluster and in the surrounding graphene. This connects with the experimental observation that the cluster can appear either brighter or darker than its environment depending on bias \cite{CranneyEPL2010,NairPRB2012}. The sign of the contrast changes more than once between the lower and upper VHS, indicating that the cluster should be understood as an energy-dependent scatterer rather than as a purely topographic object.

The same $T$ matrix introduced in the momentum-space QPI calculation can be
used to compute the real-space LDOS modulation. The only difference is the
basis in which the external propagation is written. In the QPI calculation, the
external states are Bloch states labelled by $\kk$ and $\kk+\qq$. In the
real-space calculation, the external point is instead a carbon atom at position
$\rr$. The internal scattering problem is unchanged: the Au impurity is still
represented by a six-site ring potential acting on the six carbon atoms
surrounding a hollow site.

For a single hollow-site impurity, the six ring sites are denoted by
$\rr_i$, with $i=1,\ldots,6$. In the six-site ring basis, the full
multiple-scattering matrix is the same object as in Eq.~\eqref{eq:ring_T}.
 We denote by
$T_{ij}(\ww)=\langle i|T(\ww)|j\rangle$ the matrix elements of this
six-site $T$ matrix. With this convention, $j$ is the incoming ring site and
$i$ is the outgoing ring site. Thus $T_{ij}(\ww)$ is not a bare hopping or a
bare potential, but the full amplitude for an electron entering the impurity
region on site $j$ to leave it on site $i$ after any number of intermediate
propagation and scattering events inside the ring.

The correction to the real-space Green's function at an observation point
$\rr$ is then obtained by propagating from $\rr$ to the ring, scattering through
$T$, and propagating back to $\rr$:

\begin{equation}
  \delta G(\rr,\rr;\ww)
  =
  \sum_{i,j=1}^{6}
  G_0(\rr,\rr_i;\ww)\,
  T_{ij}(\ww)\,
  G_0(\rr_j,\rr;\ww).
  \label{eq:dG_realspace}
\end{equation}
Here $G_0(\rr,\rr_i;\ww)$ describes propagation in pristine graphene from the
observation point to carbon atom $i$ of the ring, $T_{ij}(\ww)$ describes the
full scattering process within the Au-induced six-site impurity complex defined
in Eq.~\eqref{eq:himp}, and $G_0(\rr_j,\rr;\ww)$ describes the
return propagation from carbon atom $j$ back to the observation point. This is
the real-space counterpart of the QPI formula: rather than resolving the
transferred momentum $\qq$, it gives the spatial standing-wave pattern
generated by the same impurity scattering process.

The LDOS correction is obtained from the imaginary part of the local Green's
function,
\begin{equation}
  \delta\rho(\rr,\ww)
  =
  -\frac{1}{\pi}
  {\rm Im}\,
  \Tr_{\rm sub}
  \delta G(\rr,\rr;\ww).
  \label{eq:drho_realspace}
\end{equation}
Since the numerical LDOS is evaluated on individual graphene carbon atoms, the
trace reduces in practice to the diagonal component corresponding to the
sublattice of the observation site.

To mimic the small clusters observed experimentally, we also consider a
three-impurity configuration. The three Au atoms are placed on hollow sites whose relative separations correspond to the graphene-lattice vector $(2,1)$ and a distance of $\sqrt{7}\,a$ between neighbouring impurity centers. This is the smallest compact three-site arrangement compatible with the cluster separations most often identified in the STM images. It gives a triangular cluster of three equivalent hollow-site ring impurities. In the numerical implementation used for the
figures, the LDOS modulation of this cluster is approximated by the sum of the
three single-ring responses,
\begin{equation}
  \delta\rho_{\rm cl}(\rr,\ww)
  =
  \sum_{a=1}^{3}
  \delta\rho_{\rm ring}(\rr-\RR_a,\ww),
  \label{eq:cluster_ldos_sum}
\end{equation}
where $\RR_a$ are the three hollow-site centers. This keeps the full six-site form factor and full intra-ring $T$ matrix for each Au atom, but neglects additional multiple scattering between different Au atoms in the cluster. This approximation isolates the effect of the ring form factor and of the three-center geometry; a fully coherent treatment would require an enlarged $18\times18$ $T$-matrix coupling all six ring sites of all three impurities: we have checked that these coherent effects are indeed negligible for this configuration. 

In the real-space figures, $\delta\rho(\rr,\ww)$ is first evaluated on the
graphene carbon sites. The displayed STM-like image is then obtained by
broadening each carbon-site value with a narrow Gaussian, with
standard deviation $\sigma=0.3a$, where $a$ is the graphene lattice
constant.

We note that the real-space calculation does not produce clearly visible standing waves with the periodicity discussed in the previous section. This is because the calculated LDOS is dominated by small-q scattering processes, whereas the contributions associated with wavevectors near the M points are comparatively weak. Their real-space signature is therefore masked and becomes apparent only after filtering the Fourier-space signal around the M regions. In the experiment, by contrast, the M-centered contributions are particularly pronounced, indicating a substantial enhancement relative to the small-q response. This enhancement may result from coherent multiple scattering between Au clusters, which could suppress or dephase the long-wavelength, small-$q$ contribution while preserving the characteristic $M$-centered scattering processes. Alternatively, it may have a different physical origin, whose identification lies beyond the scope of the present work.

\begin{figure*}[t]
  \centering
  \vspace{-1in}
  \includegraphics[width=1\textwidth]{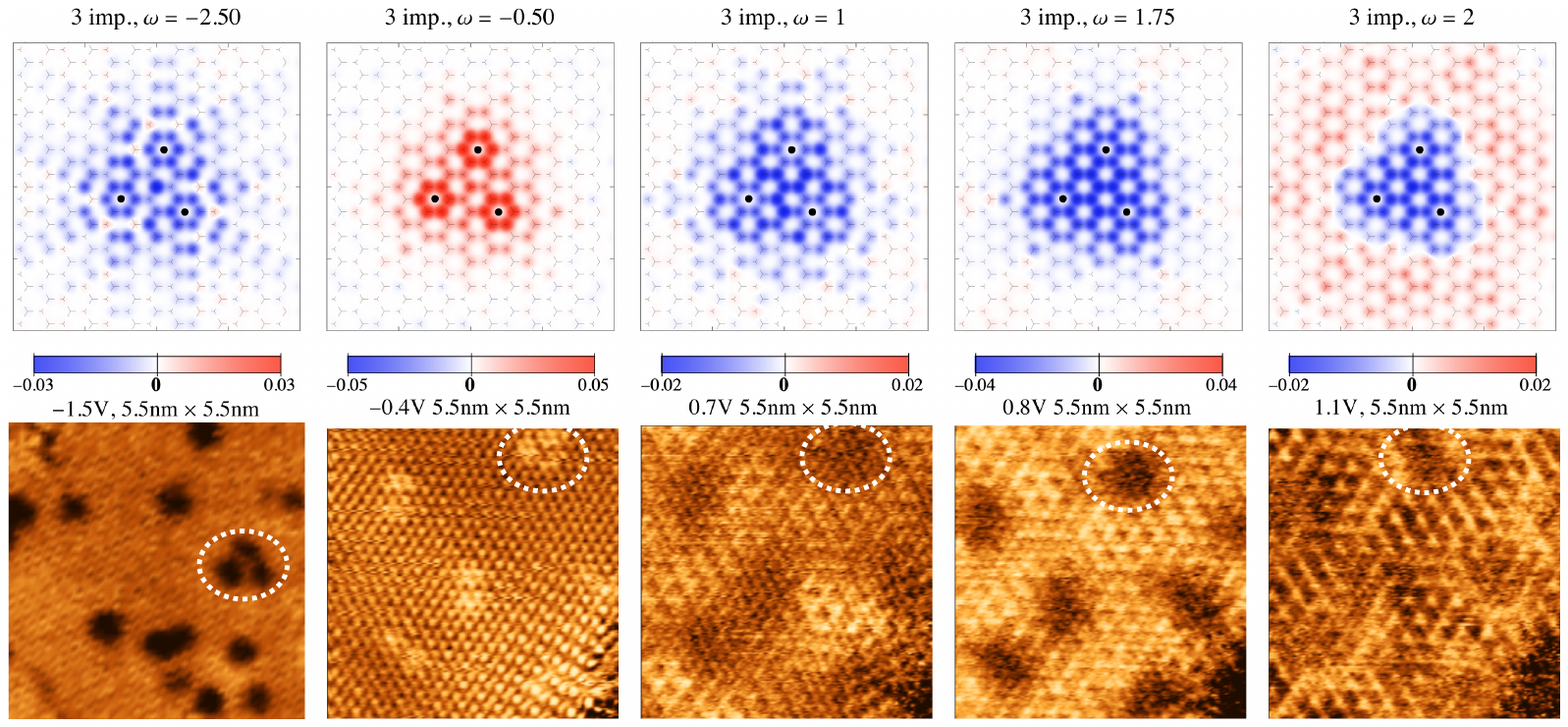}
      \vspace{-1.2in}
  \caption{Top row: calculated LDOS correction for a representative three-atom cluster at selected energies (in arbitrary units), using $U=$ 3 eV. Blue and red indicate negative and positive corrections, corresponding respectively to darker and brighter regions relative to the experimental background.  Lower row: experimental plots showing the dependence of the LDOS with position at various energies. Except for $E=$ -1.5eV, the panels describe approximatively the same spatial region on which we can identify roughly 5 clusters. High-intensity regions appear light, whereas low-intensity regions appear dark. A representative three-atom cluster is outlined by the dashed contour. The strongly energy-dependent contrast in the theoretical calcuations, including multiple sign changes, is in qualitative agreement with experiment.}
  \label{fig:cluster_ldos_maps}
\end{figure*}

In the top row of Figure~\ref{fig:cluster_ldos_maps} we plot the calculated $\delta\rho_{\rm cl}(\rr,\ww)$ at several energies. Close to the lower VHS, the impurity gives rise to a negative LDOS correction relative to the background and appears darker in maps sensitive to this energy window. At less negative energies, the sign of the LDOS correction becomes positive. Around the Dirac point and toward the upper VHS, further sign changes can occur: at small positive energies the LDOS correction is negative, and the contrast becomes stronger as one approaches the upper VHS. Once the energy moves beyond this region, both the average LDOS correction and the cluster--background contrast are strongly reduced. 

In the bottom row of the same figure we give some representative plots of the measured LDOS at various energies. It is impossible to make a quantitative comparison between the energies considered in the theory and the experiment because of Au- and substrate-induced modifications of the experimental graphene band structure, including uncertainty in the exact position of the Dirac point and in the energies of the VHSs. However, one can see that there is a qualitative agreement in the trend of the LDOS features: sharp negative contrast at very large negative energies, a positive contrast at smaller negative energies, a negative contrast at positive energies that increases with increasing energy, and then fades out at very large energies. Moreover the gold atoms give rise to localized features such that the threefold structure of the clusters is visible at negative energies, while at positive energies the features are more spread, consistent with the theoretical results.

\begin{figure}[t]
  \centering
  \includegraphics[width=0.6\textwidth, angle=270]{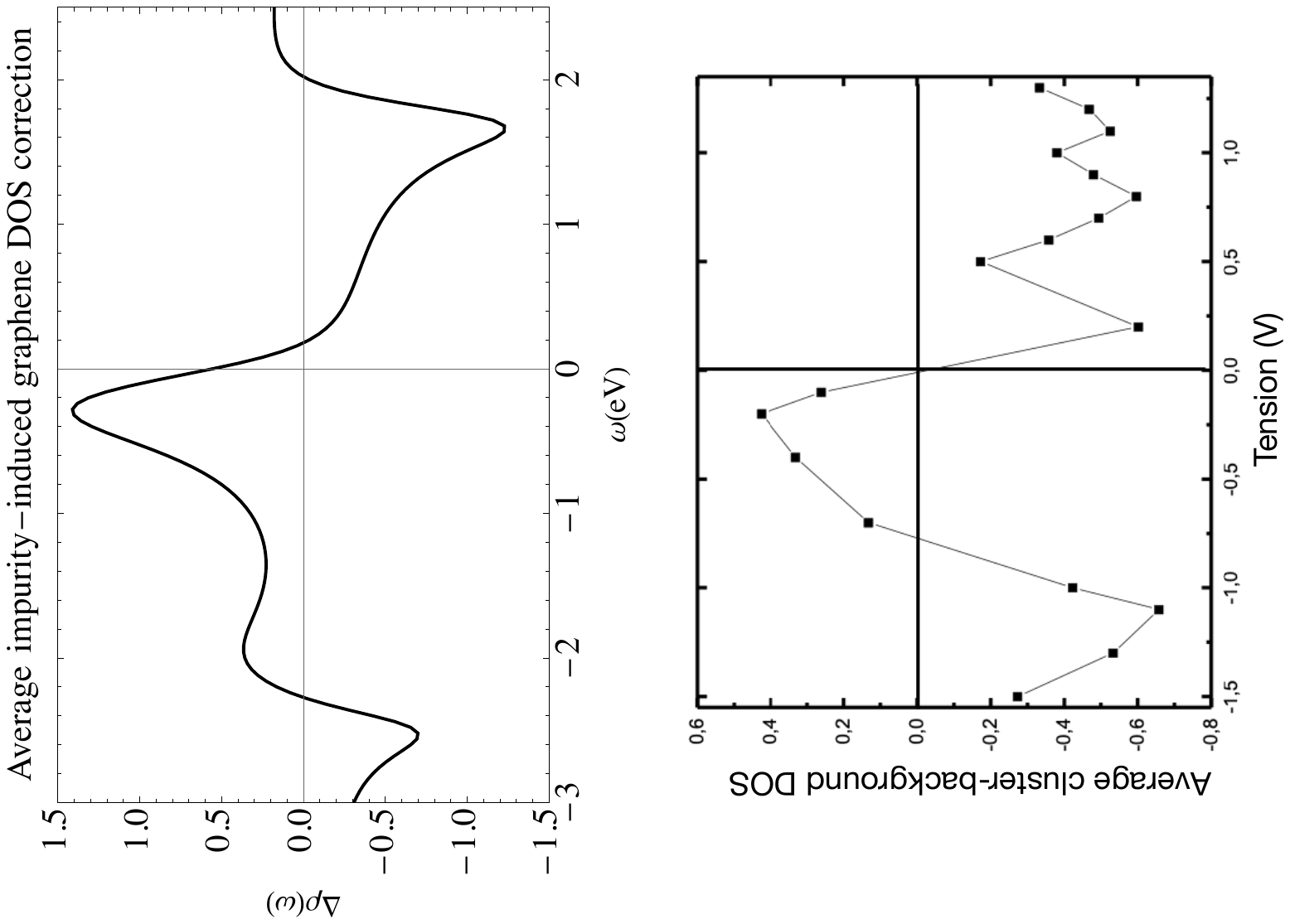}
  \caption{DOS variation as a function of energy. Upper panel: theoretical calculations of the average DOS modification (in arbitrary units), we take $U=$ 3eV. Lower panel: experimental cluster-minus-background signal, normalized to the background. Note the presence of several sign changes, in both the theoretical calculations and the experimental measurements.}
  \label{fig:dos_modification}
\end{figure}

This behaviour is summarized by the spatially averaged DOS modification shown in Fig.~\ref{fig:dos_modification}. The DOS correction changes sign multiple times, consistently with the contrast inversions observed experimentally. 

\section{Conclusions}
\label{sec:concl}

We have proposed a microscopic explanation for the unusual QPI and standing-wave features observed in the Au-cluster phase of graphene on SiC. The central mechanism is the scattering of graphene electrons by a hollow-site, ring-like impurity. When the QPI is computed using the corresponding $T$ matrix, the Fourier map develops compact double-elliptic structures centered near the graphene $M$ points, matching the main experimental signature of the cluster phase. A real-space Fourier analysis of these $M$-centered structures shows that they naturally generate standing waves with wavevectors that are nearly energy independent. In addition, a direct real-space calculation for a small cluster shows that the LDOS contrast on and off the cluster is strongly energy dependent and can change sign multiple times, consistent with experimental observations.

\begin{acknowledgments}
We thank Igor de Melo Froldi, Adeline Cr\'epieux, Catherine P\'epin, and Emile Pangburn for useful discussions and comments.
This work was supported by the ``Action amor\c{c}age du programme recherche \`a risque CEA - Audace!'' project ``Flat bands and high-temperature superconductivity in graphene''.
\end{acknowledgments}

\end{document}